\documentclass[times,twocolumn,review]{elsarticle}

\usepackage{jcomp}
\usepackage{framed,multirow}

\usepackage{amssymb}
\usepackage{latexsym}

\usepackage{url}
\usepackage{xcolor}
\definecolor{newcolor}{rgb}{.8,.349,.1}

\usepackage{amsmath}
\usepackage{subcaption}
\graphicspath{{./img/}}

\journal{Journal of Computational Physics}

\begin{document}

\verso{Sina Ackermann \textit{etal}}

\begin{frontmatter}

\title{Multi-scale three-domain approach for coupling free flow and flow in porous media including droplet-related interface processes}

\author[1]{Sina \snm{Ackermann}\corref{cor1}}
\cortext[cor1]{Corresponding author:
  sina.ackermann@iws.uni-stuttgart.de}
\author[1]{Carina \snm{Bringedal}}
\author[1]{Rainer \snm{Helmig}}

\address[1]{Institute for Modelling Hydraulic and Environmnetal Systems, Pfaffenwaldring 61, 70569 Stuttgart, Germany}

\received{\today}
\finalform{\today}
\accepted{\today}
\availableonline{\today}
\communicated{\today}

\begin{abstract}
Drops on a free-flow/porous-medium-flow interface have a strong influence on the exchange of mass, momentum and energy between the two macroscopic flow regimes.
Modeling droplet-related pore-scale processes in a macro-scale context is challenging due to the scale gap,
but might be rewarding due to relatively low computational costs.
We develop a three-domain approach to model drop formation, growth, detachment and film flow in a lower-dimensional interface domain.
A simple upscaling technique allows to compute the drop-covered interface area fraction which affects the coupling fluxes.
In a first scenario, only drop formation, growth and detachment are taken into account.
Then, spreading and merging due to lateral fluxes are considered as well.
The simulation results show that the impact of these droplet-related processes can be captured. However, extensions are necessary to represent the influence on the free flow more precisely.
\end{abstract}

\begin{keyword}
Drops, coupling, lower-dimensional domain, porous media, interfaces
\end{keyword}

\end{frontmatter}


\section{Introduction}
\subsection{Motivation}
Drop formation at the interface between a single-phase gaseous free flow and a two-phase flow in a hydrophobic porous medium occurs in many technical applications.
Examples are the water management in fuel cells, thermal insulation of building exteriors or turbine heat exchange processes.
With an appropriate pressure gradient, the liquid phase in the porous medium flows towards the interface where it enters the surface pores.
For liquid water in a hydrophobic porous medium,
drops form on the pore throats and grow into the adjacent free-flow domain.
Depending on the surrounding flow conditions, the drops might spread and merge, or be detached by the free flow.

In fuel cells, for example, the drops might block the surface and therefore prevent the
reaction of air and oxygen.
To describe the processes happening in such applications, two spatial scales are distinguished.
On the pore-scale, detailed information about pore sizes, individual drop volumes and drop surfaces is given.
The distribution of liquid and gas and their respective interfaces can be described exactly.
Averaging all fluid and material properties over representative elementary volumes (REVs)
allows a description on the macro-scale, which is usually sufficiently precise for real-life scenarios.
On the macro-scale, phase interfaces are no longer resolved.

In order to predict the consequences of drops on coupled free-flow/porous-medium-flow systems, numerical simulators are developed, as for example in \cite{Berning2009} and \cite{Baber2016}.
In these cases, the coupled systems of free flow and flow in porous media are modeled with the help of macro-scale concepts.
Even though these models do not take detailed pore-scale information into account, the results are often precise enough for real-world scenarios.
In contrast, droplet-related processes are usually studied with pore-scale models which consider properties such as interphasial areas and varying contact angles.
Drop dynamics and droplet-related flow processes differ from the multi-phase flow patterns which are assumed and described on the macro-scale.
Therefore, the pore-scale drop dynamics need to be combined with the macro-scale model for flow and transport processes in coupled free-flow/porous-medium-flow systems.

The aim of this work is to obtain a model concept on the REV-scale that includes pore-scale droplet-related processes.
We therefore develop a multi-scale model which contains droplet-related pore-scale information.
The upscaling procedure is designed in such a way that the properties which influence the exchange of mass, momentum and energy are preserved in the macro-scale description.

\subsection{State of the art}
In the following literature review, we first refer to models for single-phase coupled free-flow/porous-medium-flow systems.
Then, approaches for coupled multi-phase systems without and with the influence of drops are presented.

For single-phase systems, commonly either a one-domain or a two-domain approach is applied.
In the one-domain approach, one set of balance equations describes the flow and transport processes in the whole system.
In the Brinkman equation derived in \cite{Brinkman1949}, the Stokes equation is combined with Darcy's law to obtain one momentum balance equation which is valid in the whole domain.
Spatial parameters such as porosity or permeability
are defined such that they represent the porous medium or the free flow domain respectively, with a smooth transition zone in between.

In the two-domain approach, two different sets of balance equations describe the respective flow regimes.
At the sharp interface in between, coupling conditions determine the exchange of mass, momentum and energy between the two domains.
Such conditions for single-phase systems are analyzed in \cite{Discacciati2009} and \cite{Riviere2005}.

For multi-phase flows in porous media, modeling the behavior of the liquid phase at the
coupling interface is challenging.
The approach for compositional non-isothermal systems presented in \cite{Mosthaf2011} and \cite{Baber2012} is based on the assumption that the normal water flux coming from the porous medium evaporates directly into the gaseous free flow when it reaches the interface. 
The same is assumed in \cite{Masson2016} and \cite{Chen2014}.
Another possibility is to assume that the liquid phase does not reach the interface and can therefore be neglected in the coupling conditions, as implemented in \cite{Dumux3}.
Both assumptions neglect the fact that liquid drops might form and move on the porous surface in such a set-up.

A multifluid approach is applied to model the accumulation of liquid water in the gas channel of a fuel cell in \cite{Berning2009}.
At the interface, the liquid pressure is set as the capillary pressure between the liquid phase pressure in the porous medium and the pore entry pressure which is derived from the Hagen-Poiseuille equation.
Another multiphase, multifluid model is presented in \cite{Gurau2008}, where pending drops in the gas channel are investigated with the help of separate transport equations for each phase.
In the concept presented in \cite{Baber2016}, the drops are added to the coupled free-flow/porous-medium-system by applying a mortar method.
The additional degrees of freedom allow to store mass and energy in the interface.
The drop volume can then be computed as an additional primary variable and is used to predict the drops' influence in fuel cells.

Within this work, we use the model developed in \cite{Baber2016} as a base to describe the droplet-related processes.
In their approach, only a small number of drops can be computed due to stability issues. In addition, it is impossible to take lateral fluxes into account.
Therefore, we implement the principles of \cite{Baber2016} in a lower-dimensional domain approach as for example presented in \cite{Glaeser2017},
to obtain more flexibility with respect to drop dynamics.

\subsection{Outline}
In the next section, we explain the model concepts for free flow, flow in porous media and drops at the interface of these two flow regimes.
Section \ref{sec:coupling} deals with the coupling concept to describe the exchange of mass, momentum and energy.
In Section \ref{sec:numerical}, the numerical model is formulated.
Then, the results of numerical simulations are presented in Section \ref{sec:results}.
Finally, a summary and outlook are given in Section \ref{sec:conclusions}.

\section{Model concepts}
\begin{figure}
 \centering
 \includegraphics[width=0.45\textwidth]{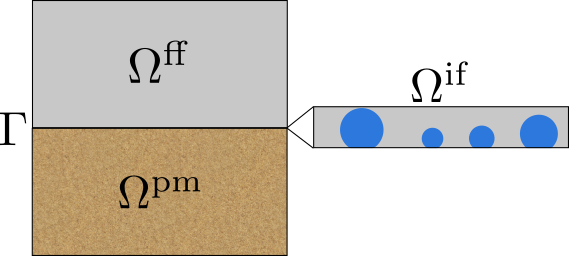}
 \caption{Three-domain approach with lower-dimensional interface domain $\Gamma$
 between the full-dimensional subdomains $\Omega^{\text{ff}}$ and $\Omega^{\text{pm}}$}
 \label{fig:three-dom}
\end{figure}

We extend the existing two-domain approach for coupled free-flow/porous-medium-flow systems \cite{Mosthaf2011}
to a three-domain approach with an additional interface domain as shown in Figure \ref{fig:three-dom}.
All droplet-related processes are computed within the interface domain, following the derivations in \cite{Baber2016}.
First, we take a full-dimensional interface domain $\Omega^{\text{if}}$ into account.
Similar to the approach in \cite{Glaeser2017}, the respective equations are then upscaled and solved in a
lower-dimensional interface domain $\Gamma$ which is defined as $\bar \Omega^{\text{ff}} \cap \bar \Omega^{\text{pm}} = \Gamma$.

We assume local thermodynamic equilibrium, i.~e. mechanical, chemical and thermal equilibrium,
and non-isothermal conditions in each domain.
The gas and liquid phases consist of the two components water and air which mix according to binary diffusion.
The mass fractions in each phase $\alpha \in \{g,l\}$ sum up to one: $X_{\alpha}^w + X_{\alpha}^a = 1$.
Further assumptions and the respective equations to be solved are given in the next three sections.
The nomenclature is listed in Table \ref{tab:nomenclature}.

\subsection{Free flow}
\label{sec:ff}
We consider a single-phase gaseous free flow under laminar conditions in $\Omega^{\text{ff}}$.
The gas is assumed to be compressible.
The total mass balance is given as
\begin{equation}
 \frac{\partial \varrho_g}{\partial t} + \nabla \cdot (\varrho_g \textbf{v}_g) = q_g\;.
 \label{eq:ff-mass}
\end{equation}
For the mass of each component $\kappa \in \{a,w\}$, the following balance holds:
\begin{equation}
 \frac{\partial (\varrho_g X_g^{\kappa})}{\partial t}
 + \nabla \cdot \left( \varrho_g \textbf{v}_g X_g^{\kappa}
- D_g^{\kappa} \varrho_g \nabla X_g^{\kappa} \right)
= q_g^{\kappa}\;.
\label{eq:ff-comp}
\end{equation}
We solve the two component mass balance equations (\ref{eq:ff-comp}) and use the supplementary relation $X_g^w + X_g^a = 1$.

The Navier-Stokes equations describe the momentum balance:
\begin{align}
\nonumber
& \frac{\partial (\varrho_g \textbf{v}_g)}{\partial t} + \nabla \cdot \left(\varrho_g \textbf{v}_g \textbf{v}_g^{\text{T}} \right) \\
&= \nabla \cdot \left( \mu_g \left(\nabla \textbf{v}_g + \nabla \textbf{v}_g^T \right) + \left(\frac{2}{3} \mu_g \nabla \cdot \textbf{v}_g \right) \textbf{I} - p_g \textbf{I} \right) + \varrho_g \textbf{g}\;.\label{eq:ff-NS}
\end{align}
The dilatation term $\nabla \cdot \left( \left( \frac{2}{3} \mu_g \nabla \cdot \textbf{v}_g \right) \textbf{I} \right)$ will be neglected in the following.
The energy balance is
\begin{equation}
  \frac{\partial (\varrho_g u_g)}{\partial t}
 + \nabla \cdot \left( \varrho_g \textbf{v}_g h_g
- \lambda_g \nabla T \right)
= q_T\;.
\end{equation}

Solving these four equations with respective initial and boundary conditions
yields the solution for the gas pressure $p_g$, the gas velocity $\textbf{v}_g$,
the mass fraction of water in gas $X_g^w$ and the temperature $T$.
To close the system of equations, additional supplementary equations are necessary, as listed in Table \ref{tab:fluid-props}.

\subsection{Flow in porous media}
\label{sec:pm}
A rigid solid phase and a two-phase flow interact with each other in the porous medium $\Omega^{\text{pm}}$.
Both fluids are assumed to behave like Newtonian fluids, the gas phase is compressible while the liqud phase is incompressible.
Instead of resolving the pore network in detail, the porous medium properties are averaged over representative elementary volumes (REVs).
The flow velocity is assumed to be small ($Re < 1$) such that Darcy's law can be applied
to compute the phase velocities:
\begin{equation}
 \textbf{v}_{\alpha} = - \textbf{K} \frac{k_{r \alpha}}{\mu_{\alpha}}
\left( \nabla p_{\alpha} - \varrho_g \textbf{g}
\right)\;.
\end{equation}

For the total mass balance, the contributions from both phases are summed up:
\begin{equation}
 \sum\limits_{\alpha \in \{l,g\}} \Phi \frac{\partial (\varrho_{\alpha} S_{\alpha})}{\partial t}
+ \nabla \cdot \left( \sum\limits_{\alpha \in \{l,g\}} \varrho_{\alpha} \textbf{v}_{\alpha} \right)
= \sum\limits_{\alpha \in \{l,g\}} q_{\alpha} \;.
\label{eq:pm-mass}
\end{equation}

The component mass balances for $\kappa \in \{a,w\}$ are given as
\begin{equation}
 \sum\limits_{\alpha \in \{l,g\}} \Phi \frac{\partial (\varrho_{\alpha} X_{\alpha}^{\kappa} S_{\alpha})}{\partial t}
 + \nabla \cdot \textbf{F}^{\kappa}
 =  \sum\limits_{\alpha \in \{l,g\}} q_{\alpha}^{\kappa}
 \label{eq:pm-comp}
\end{equation}
with the flux term
\begin{equation}
 \textbf{F}^{\kappa} = \sum\limits_{\alpha \in \{l,g\}} (\varrho_{\alpha} \textbf{v}_{\alpha} X_{\alpha}^{\kappa}
 - D_{\alpha}^{\kappa,pm} \varrho_{\alpha} \nabla X_{\alpha}^{\kappa}) \;.
\end{equation}
Due to the assumption of local thermodynamic equilibrium, all present phases have the same temperature: $T_g = T_l = T_s = T$.
Therefore, we can formulate the energy balance as
\begin{align}
\nonumber
 &\sum\limits_{\alpha \in \{l,g\}} \Phi \frac{\partial (\varrho_{\alpha} u_{\alpha} S_{\alpha})}{\partial t}
+ (1-\Phi) \frac{\partial (\varrho_s c_s T)}{\partial t}\\
&+ \nabla \cdot \left( \sum\limits_{\alpha \in \{l,g\}}
\varrho_{\alpha} \textbf{v}_{\alpha} h_{\alpha} - \lambda_{\text{pm}} \nabla T \right)
= q_T\;.
\end{align}

In addition to the balance equations, $S_g + S_l = 1$ holds and the relationship between the phase pressures is given as $p_c(S_w) = p_n - p_w$,
where the capillary pressure $p_c$ depends on the wetting-phase saturation $S_w$.
With these relationships, the gas pressure $p_g$, the liquid-phase saturation $S_l$ and the temperature $T$
can be obtained by solving the balance equations with respective initial and boundary conditions.
If the liquid phase disappears due to evaporation, the mass fraction of water in gas $X_g^w$ becomes the new primary variable instead of the saturation $S_l$ (\cite{Class2002}).
The remaining variables are either material parameters or can be computed with additional supplementary relations. Again, we refer to Table \ref{tab:fluid-props} for details.
The capillary pressure $p_c(S_w)$ and the relative permeabilities $k_{r \alpha}$ are computed with a regularized van-Genuchten model (\cite{vanGenuchten1980}).

\begin{table}
 \centering
 \begin{tabular}{l c}
 $\varrho_g = \varrho_g^a + \varrho_g^w = \frac{p_g M^a}{RT}
  + \varrho_g^w(T,p)$ & ideal gas, \\
 & \cite{iapws1997} \\[1ex]
 & \\[-2ex]
 $\varrho_l = \varrho_l^{\text{mol},w} (M^w x_l^w + M^a x_l^a)$ & \cite{Class2002} \\[1ex]
 & \\[-2ex]
 $D_g = 2.13 \cdot 10^{-5} ~ \frac{10^5}{p_g} ~ \left(\frac{T}{273.15}\right)^{1.8} $ & \cite{Vargaftik1975}\\[1ex]
 & \\[-2ex]
 $D_l = 2.01 \cdot 10^{-9} \frac{T}{273.15 + 25}$ & \cite{Reid1987}\\[1ex]
 & \\[-2ex]
 $h_g = X_g^w h^w + X_g^a h^a$ & \cite{Class2002}\\[1ex]
 & \\[-2ex]
 $h_l = h_l^w$ & \\[1ex]
 & \\[-2ex]
 $h_g^a = c_g^a(T - 273.15)$ &  \cite{Kays2005} \\[1ex]
 & \\[-2ex]
 $h_{\alpha}^w = h_{\alpha}^w(p_{\alpha},T)$ & \cite{iapws1997} \\[1ex]
 & \\[-2ex]
 $c_g = c_g^a x_g^a + c_g^w x_g^w$ & \\[1ex]
 & \\[-2ex]
 $c_l = c_l^w$ &  \\[1ex]
 & \\[-2ex]
 $c_g^a = c_g^a(T,p_g)$ & \cite{Hollis1996} \\[1ex]
 & \\[-2ex]
 $c_{\alpha}^w = c_{\alpha}^w (T, p_{\alpha})$ & \cite{iapws1997} \\[1ex]
 & \\[-2ex]
 $ u_{\alpha} = h_{\alpha} - \frac{p_{\alpha}}{\varrho_{\alpha}}$ & \\[1ex]
 & \\[-2ex]
 $\gamma_{lg} = 0.2358 \cdot (1-\frac{T}{T_c})^{1.256}$ & \\[1ex]
 & \\[-2ex]
 \hspace{4em} $\cdot \left(1 - 0.625(1-\frac{T}{T_c}) \right)$ & \cite{iapws1994} \\[1ex]
 & \\[-2ex]
 $\lambda_g = \lambda_g^a = 0.0255535$ & \cite{NIST} \\[1ex]
 & \\[-2ex]
 $\lambda_l = \lambda_l^w(T,p_l)$ & \cite{iapws2011} \\[1ex]
 & \\[-2ex]
 $\mu_g$ according to the Wilke method & \cite{Reid1987} \\[1ex]
 & \\[-2ex]
 $\mu_l = \mu_l^w$ &
 \end{tabular}
 \caption{Fluid properties for the phases and components used in this work}
 \label{tab:fluid-props}
\end{table}

\subsection{Interface}
\label{sec:if}
The interface domain $\Omega^{\text{if}}$ consists of a thin layer of the free-flow region  as shown on the right side in Figure \ref{fig:three-dom}.
We assume the top layer of the porous medium to consist of parallel, circular and
perpendicular pores.
These pores allow mass and energy fluxes between porous medium and interface.
On the upper boundary of the interface domain, a smooth transition into the free-flow
domain is given.
In contrast to the balance equations in the two previous sections, the influence of pore-scale drops is included in the REV-scale equations for the interface domain.
The transition from a full-dimensional description for $\Omega^{\text{if}}$ towards a lower-dimensional model concept for $\Gamma$ is outlined in the following.

\subsubsection{Assumptions}
\begin{figure}
 \begin{subfigure}{0.5\textwidth}
  \includegraphics[width=\linewidth]{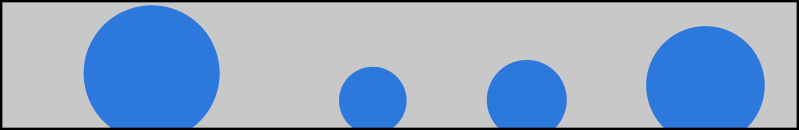}
  \caption{Scenario 1: Formation, growth, detachment}
  \label{fig:detachment-domain}
 \end{subfigure}
 \begin{subfigure}{0.5\textwidth}
  \includegraphics[width=\linewidth]{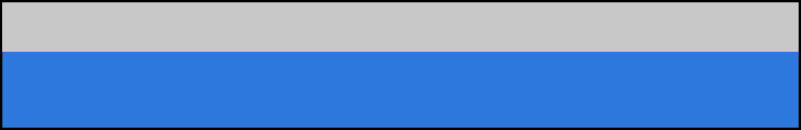}
  \caption{Scenario 2: Formation, growth, film flow}
  \label{fig:filmflow-domain}
 \end{subfigure}
 \caption{Schematic view of the two scenarios for drops at the interface of a free flow and a porous medium}
 \label{fig:scenarios}
\end{figure}

We assume two different scenarios in the scope of this work, as illustrated in Figure \ref{fig:scenarios}.
In the first scenario, the drops grow and detach, but do not interact.
In the second scenario, drops can touch and spread along the surface.

For the first scenario (Figure \ref{fig:detachment-domain}), we assume the drops to be symmetrical with a circular contact line.
They do not deform due to shear forces exerted from the surrounding flow field.
Horizontal fluxes along the interface are neglected,
leading to constant contact angles but varying contact areas due to drop growth.
With these assumptions, drop formation, growth and detachment can be modeled.
Detached drops are assumed to be transported away with the free flow and do not interact with the interface anymore.

The second scenario (Figure \ref{fig:filmflow-domain}) takes spreading and merging into account.
If the drops merge, film flow can occur along the interface.
Lateral fluxes of both the gas and the liquid phase are modeled, while the sliding, rolling and break-up of drops is still neglected.

\subsubsection{Transfer from pore- to REV-scale}
\label{sec:homog}
The following concept is based on the simple upscaling and homogenization technique presented in \cite{Baber2016}.
Considerations on the pore-scale with the help of a bundle-of-tubes model lead to an REV-scale concept which allows to take droplet-related processes into account.
In most porous media, not all pores have the same radius.
By clustering them in $N$ pore classes with a respective mean pore radius $\bar{r}_{\text{pore}}$, condition (\ref{eq:formation}) has to be evaluated only $N$ times.
The fate of the drops in each pore class can be determined by evaluating the liquid fluxes into the drops and the balance of drag and retention forces for each pore radius separately.
Then, the drop volumes of the individual drops are summed up as
\begin{equation}
 V_{\text{drop}}^{\text{sum}} = \sum_{i \in \{1, \dots, N\}} V_{\bar{r}_{\text{pore},i}} n_{\bar{r}_{\text{pore},i}},
\end{equation}
where $n_{\bar{r}_{\text{pore},i}}$ is the number of pores in the respective pore class.

\begin{figure}
 \centering
 \includegraphics[width=0.5\textwidth]{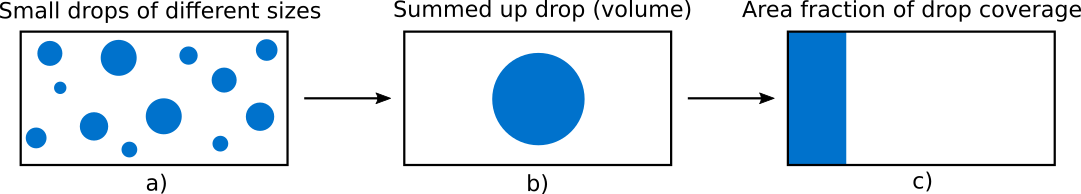}
 \caption{Upscaling: The individual drop volumes are summed up to $V_{\text{drop}}^{\text{up}}$, which is used to compute the drop-covered area
 fraction $a_{\text{drop}}$.}
 \label{fig:upscaling}
\end{figure}

In the following, we determine the area fractions of the interface $\Gamma$ that are available for mass fluxes between the interface and its neighboring compartments.
Summing up the cross-sectional areas of all liquid-filled pores yields the total liquid-filled area $A_l^{\text{if}}$.
The area fraction available for liquid fluxes between porous medium and interface is then
given as
\begin{equation}
 a_l = \frac{A_l^{\text{if}}}{\Phi A_{\Gamma}}\;.
 \label{eq:al}
\end{equation}
This procedure is also illustrated in Figure \ref{fig:upscaling}.
For the gas flux between porous medium and interface, the remaining area fraction is
\begin{equation}
 a_g = 1 - a_l \;.
 \label{eq:agpm}
\end{equation}
The area fractions will be used later to compute the fluxes across the interface.

\subsubsection{Lower-dimensional interface domain}
With the upscaling procedure explained in the previous section, a lower-dimensional model concept can be formulated for the interface domain.
The idea to model the droplet-related processes in a lower-dimensional interface domain
is inspired by the approach in \cite{Glaeser2017}, where fractures in porous media
are represented by lower-dimensional domains.
To conserve the quantities in the interface domain, the general balance equations have to be adapted to take the drops' influence into account.

In the presented approach, the drops influence only the area for the exchange
of mass and energy between the free flow and the flow in the porous medium.
Due to the small size of the drops, we assume that their influence on the free-flow velocity field can be neglected.
The drop volumes are computed as secondary variables.
The corresponding water volume in the interface domain is taken into account in the saturation $S_l$, which is a primary variable.

Total mass conservation is given by a balance of the mass storage, fluxes and sinks:
\begin{equation}
 \sum\limits_{\alpha \in \{l,g\}} \frac{\partial (\varrho_{\alpha} S_{\alpha})}{\partial t}
 + \nabla \cdot \sum\limits_{\alpha \in \{l,g\}} \textbf{F}_{\alpha}
 = q^{\text{ff}} + q^{\text{pm}} - q^{\text{detach}}\;.
 \label{eq:if-mass}
\end{equation}
The fluxes between the flow regimes have to be included in the
sink and source terms $q^{\text{ff}}$ and $q^{\text{pm}}$.
These terms will be defined in the drop coupling concept in Section \ref{sec:dropcoupling}.
When a drop detaches, its mass is removed from the system, which is modeled with the additional sink term
$q^{\text{detach}}$:
\begin{equation}
 q^{\text{detach}} =
 \begin{cases}
0 & \text{drop has not detached} \\
\frac{\varrho_l V_{\text{drop}}}{\Delta t} & \, \text{drop has detached.}
\end{cases} \label{eq:qif}
\end{equation}

The same principle applies to the component mass balances:
\begin{eqnarray}
&& \sum\limits_{\alpha \in \{l,g\}} \dfrac{\partial (\varrho_{\alpha} X_{\alpha}^{\kappa} S_{\alpha})}{\partial t} + \nabla \cdot \sum\limits_{\alpha \in \{l,g\}}
\textbf{F}_{\alpha}^{\kappa} \nonumber \\
&&= q^{\text{ff},\kappa} + q^{\text{pm},\kappa} - q^{\text{detach},\kappa}\;.
 \label{eq:if-comp}
\end{eqnarray}
Again, the source terms represent the coupling fluxes and the mass loss due to detached drops.

For the energy balance equation, the source terms are formulated accordingly:
\begin{eqnarray}
 &&\sum\limits_{\alpha \in \{l,g\}}  \dfrac{\partial (\varrho_{\alpha} u_{\alpha} S_{\alpha})}{\partial t}\nonumber + \nabla \cdot \sum\limits_{\alpha \in \{l,g\}} \textbf{F}_{\alpha}^T \nonumber \\
&&= q^{\text{ff},T} + q^{\text{pm},T} - q^{\text{detach},T}\;.
 \label{eq:if-energy}
\end{eqnarray}
With this approach, the interface can now store mass and energy in the drop volumes.
This means, the interface has thermodynamic properties in contrast to the model concept presented in \cite{Mosthaf2011}.

In a first scenario, only vertical fluxes are assumed.
Therefore, the flux terms $\textbf{F}_{\alpha}$, $\textbf{F}_{\alpha}^{\kappa}$ and $\textbf{F}_{\alpha}^T$ for $\alpha \in \{g,l\}$ are zero.
Consequently, mass, momentum and energy are only exchanged with the free flow and the porous medium, not along the interface domain.
Drops can only form, grow and detach, but not move along the interface or merge.

\subsubsection{Drop formation, growth and detachment}
\label{subsec:drops}
Water drops form at the surface of a hydrophobic porous medium due to a pressure gradient
towards the porous surface.
If the liquid pressure $p_l^{\text{pm}}$ is larger than a surface pore's entry pressure $p_e$ and the gas pressure $p_g^{\text{ff}}$ above the pore, a drop forms:
\begin{equation}
 p_l^{\text{pm}} \ge \underbrace{-\frac{2 \gamma_{lg} \cos \theta}{r_{\text{pore}}}}_{p_e} + p_g^{\text{ff}}\;.
 \label{eq:formation}
\end{equation}
Since $\theta > 90^{\circ}$, $\cos (\theta)$ is negative and $p_e$ is multiplied by $-1$ to obtain the total pressure which has to be exceeded by $p_l^{\text{pm}}$.

Once formed, the drops are fed by liquid fluxes from the porous medium due to the pressure gradient towards the interface.
For $\textbf{v}_g^{\text{ff}} > 0$, the surrounding free flow pulls on the drops and exerts a drag force
\begin{equation}
 F_{\text{drag}} = \frac{1}{2} \varrho_g v_{g,t}^2 c_d (Re) A_{\text{proj}}\;,
\end{equation}
where $A_{\text{proj}}$ is the projected drop area perpendicular to the flow direction.
The drag coefficient is defined as
\begin{equation}
 c_d = 46.247 \left( \frac{d_{\text{drop}}}{h_{\text{channel}}}\right)^{0.1757} \cdot \text{Re}^{0.2158 \frac{d_{\text{drop}}}{h_{\text{channel}}} - 0.6384} \;,
\end{equation}
as suggested in \cite{Cho2012}.
Opposed to the drag force, the retention force, holds the drop on the surface
\begin{equation}
 F_{\text{ret}} = 2 r_{\text{drop}} \pi \gamma_{lg} \sin^2 (\theta) \sin\left(\frac{\Delta \theta}{2}\right)\;,
\end{equation}
where $\Delta \theta$ is the contact angle hysteresis between advancing and receding contact angle (\cite{Cho2012}).

If the drag force exceeds the retention force, the drop detaches from the surface,
which is expressed by the detachment criterion
\begin{equation}
  F_{\text{drag}} > F_{\text{ret}}\;.
  \label{eq:detachment}
\end{equation}

A new drop might form if the formation condition (\ref{eq:formation}) is still fulfilled.
Depending on the surface tension and free-flow velocity, other dynamic processes such as spreading, merging and film flow might occur.
Detached drops can either be transported away from the surface immediately, or they can break up, roll or slide on the surface.
If the liquid flux from the porous medium and the evaporative flux into the free flow balance out, the drop volume stays constant.
Drops can also form due to condensation of water vapor on the porous surface and shrink due to evaporation or seepage into the porous medium, or both.

\subsubsection{Film flow}\label{subsec:filmflow}

For a more detailed description, we take lateral fluxes along the interface into account
in a second scenario.
The free-flow shear forces might deform the drops such that they spread on the surface and merge.
Merged drops build a thin film on the surface.
This film flow only persists on a hydrophobic surface if a continuous water flux from the porous medium is provided.

For the advective gas fluxes along the interface, the free flow velocity $\textbf{v}_g^{\text{if}}$ is reduced by the presence of the drops.
We assume the following $k_r-S_l$ relationship to account for this reduction:
\begin{equation}
 k_{rg} = (1 - S_l^2) \left(  2 (S_l + 2) -  3 \frac{\mu_l (1 - S_l^2) + \mu_g S_l^2}{\mu_l ( 1- S_l) + \mu_g S_l} \right)\;.\label{eq:krg}
 \end{equation}
This derivation for this approach is given in \ref{sec:appendix}.
The gas phase velocity can then be computed as
\begin{equation}
 \textbf{v}_g^{\text{if}} = - \textbf{K} \frac{k_{rg}}{\mu_g} (\nabla p_g - \varrho_g \textbf{g})\;,
\end{equation}
with $\textbf{K} = \frac{h^2}{12}$, yielding the gaseous mass flux
\begin{equation}
 \textbf{F}_g = (\varrho_g \textbf{v}_g^{\text{if}}) \cdot \textbf{n}^{\text{if}}\,
\end{equation}
with $\textbf{n}^{\text{if}} \perp \textbf{n}^{\text{pm}}$.

\begin{figure}
 \centering
 \includegraphics[width=0.4\textwidth]{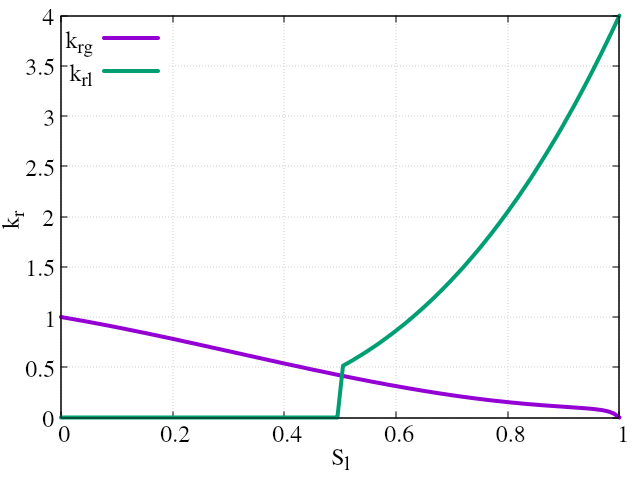}
 \caption{Exemplary $k_r-S_l$ relationships for $S_l^{\text{merge}} = 0.5$}
 \label{fig:kr-sl}
\end{figure}

If many drops have formed, the distances between them become very small such that they eventually touch and merge.
Therefore, we assume that the liquid phase velocity in the interface depends on the liquid saturation via the relative permeability $k_{rl}$:
\begin{equation}
 k_{rl} = \begin{cases}
0 & S_l \le S_l^{\text{merge}}\\
4 S_l^3 &  S_l > S_l^{\text{merge}}
\end{cases} \;.\label{eq:krl}
\end{equation}
The $k_r-S_l$ relationships are depicted in Figure \ref{fig:kr-sl}.
Note that the liquid saturation is the non-wetting phase saturation in our case.

For $S_l^{\text{if}} \ge S_l^{\text{merge}}$, film flow is assumed.
The liquid mass flux is given as
\begin{equation}
 \textbf{F}_l = \left( - \varrho_l \textbf{K} \frac{k_{rl}}{\mu_l} (\nabla p_l - \varrho_l \textbf{g})\right) \cdot \textbf{n}^{\text{if}}
  = (\varrho_l \textbf{v}_l^{\text{if}}) \cdot \textbf{n}^{\text{if}},
\end{equation}
with $\nabla p_l = \nabla p_g + \nabla p_c$.

\begin{table}
 \begin{tabular}{l l }
 Greek letters & \\
 $\gamma$ & surface tension ($\frac{N}{m}$) \\
 $\lambda$ & heat conductivity ($\frac{J}{m~s~K}$) \\
 $\theta$ & contact angle ($\deg$) \\
 $\mu$ & dymamic viscosity ($Pa~s$) \\
 $\varrho$ & mass density ($\frac{kg}{m^3}$) \\
 $\Phi$ & porosity ($-$) \\

 Latin letters & \\
 $c$ & specific heat capacity ($\frac{J}{kg~K}$) \\
 $D$ & diffusion coefficient ($\frac{m}{s}$) \\
 $\textbf{g}$  & gravity vector ($\frac{m}{s^2}$) \\
 $h$ & enthalpy ($J$) \\
 $\textbf{I}$ & identity tensor ($-$) \\
 $\textbf{K}$ & instrinsic permeability tensor ($m^2$) \\
 $k_r$ & relative permeability ($-$) \\
 $p$ & pressure ($Pa$) \\
 $q$ & source/sink term \\
 $S$ & saturation ($-$) \\
 $T$ & temperature ($K$) \\
 $t$ & time ($s$) \\
 $u$ & specific internal energy ($\frac{J}{kg}$) \\
 $\textbf{v}$ & velocity ($\frac{m}{s}$)\\
 $X$ & mass fraction ($-$)\\
 Subscripts& \\
 $\alpha$ & phase \\
 $g$ & gas phase \\
 $l$ & liquid phase \\
 $n$ & non-wetting phase \\
 $s$ & solid phase \\
 $w$ & wetting phase \\
 Superscripts& \\
 $\kappa$ & component \\
 $a$ & air \\
 $\text{ff}$ & free flow \\
 $\text{if}$ & interface domain \\
 $\text{pm}$ & porous medium \\
 $w$ & water
 \end{tabular}
\caption{Nomenclature}
\label{tab:nomenclature}
\end{table}

\section{Coupling concept}
\label{sec:coupling}
Modeling coupled flow regimes requires conditions to describe the exchange of mass, momentum and energy.
In the following, we present the established simple coupling concept for free-flow/porous-medium-flow systems first.
Then, we introduce the three-domain approach which takes interface drops into account.

\subsection{Simple coupling concept}
\label{sec:coupling:simple}
The following coupling concept for compositional, non-isothermal systems was first presented in \cite{Mosthaf2011}.
It is based on the assumption of a local thermodynamic equilibrium at the interface,
as well as the continuity of fluxes across the interface.

\subsubsection{Local thermodynamic equilibrium}
For the local mechanical equilibrium, the normal forces acting on the interface from both
sides have to be equal:
\begin{equation}
\left[ \textbf{n} \cdot \left( \left( \varrho_g \textbf{v}_g \textbf{v}_g^T
- \mu_g (\nabla \textbf{v}_g + \nabla \textbf{v}_g^\text{T})
+ p_g \textbf{I} \right) \textbf{n} \right) \right]^{\text{ff}}
= \left[ p_g \right]^{\text{pm}}\;.
\label{eq:mech-equil}
\end{equation}

We apply the Beavers-Joseph-Saffman condition \cite{Saffman1971} for the tangential forces.
This condition is actually a Robin boundary condition for the tangential velocity component
in the free flow domain:
\begin{equation}
 \left[ \left( - \frac{\sqrt{K}}{\alpha_{BJ}} (\nabla \textbf{v}_g + \nabla \textbf{v}_g^T) \textbf{n} \right) \cdot \textbf{t}_i \right]^{\text{ff}}
 = \left[ \textbf{v}_g \cdot \textbf{t}_i \right]^{\text{ff}}\;.
 \label{eq:BJS}
\end{equation}

For local chemical equilibrium, the chemical potential has to be continuous across the interface.
Due to the possible pressure jump in condition (\ref{eq:mech-equil}), the pressure ist not locally constant as stated in the definition of chemical equilibrium.
Therefore, we require the continuity of mole fractions across the interface:
\begin{equation}
 [x]^{\text{ff}} \approx [x]^{\text{pm}}\;.
 \label{eq:chem-equil}
\end{equation}

Local thermal equilibrium can be assumed by the continuity of temperature:
\begin{equation}
 [T]^{\text{ff}} \approx [T]^{\text{pm}}\;.
 \label{eq:therm-equil}
\end{equation}

\subsubsection{Continuity of fluxes}
The continuity of normal mass fluxes is given by
\begin{equation}
\left[ \left( \varrho_g \textbf{v}_g \right) \cdot \textbf{n} \right]^{\text{ff}}
= - \left[ \left( \varrho_g \textbf{v}_g + \varrho_l \textbf{v}_l \right) \cdot \textbf{n}
\right]^{\text{pm}}\;.
\label{eq:mass-flux}
\end{equation}
For the normal component mass fluxes, continuity yields
\begin{align}
&\left[ \left(
 \varrho_g X_g^{\kappa} \textbf{v}_g - D_g^{\kappa} \varrho_g
\nabla X_g^{\kappa} \right) \cdot \textbf{n} \right]^{\text{ff}} \nonumber \\
&= - \left[\sum\limits_{\alpha \in \{l,g\}} \left( \varrho_{\alpha} X_{\alpha}^{\kappa}
\textbf{v}_{\alpha} - D_{\alpha}^{\kappa, \text{pm}}
\varrho_{\alpha} \nabla X_{\alpha}^{\kappa}
\right) \cdot \textbf{n} \right]^{\text{pm}}\;.
\label{eq:comp-flux}
\end{align}
The heat flux continuity is achieved by the condition
\begin{align}
&\left[ \left(
 \varrho_g h_g \textbf{v}_g
 - \lambda_g^{\text{ff}}
\nabla T
\right) \cdot \textbf{n} \right]^{\text{ff}} \nonumber \\
&= - \left[ \left(
 \sum\limits_{\alpha \in \{l,g\}} \varrho_{\alpha} h_{\alpha}
\textbf{v}_{\alpha} - \lambda^{\text{pm}} \nabla T
 \right)
\cdot \textbf{n} \right]^{\text{pm}}\;.
\label{eq:heat-flux}
\end{align}
In Equations (\ref{eq:mass-flux}), (\ref{eq:comp-flux}) and (\ref{eq:heat-flux}), direct evaporation of the liquid phase at the coupling interface is assumed.
Details and derivations are given in \cite{Mosthaf2011}.

\subsection{Drop coupling concept}
\label{sec:dropcoupling}
In the following, the influence of the drops is taken into account with the help of an additional interface domain as described in Section \ref{sec:if}.
In contrast to the two-domain approach, two sets of coupling conditions become necessary for the three-domain approach.
Free flow and interface as well as interface and porous medium are coupled directly.
Interactions between the free flow and the flow in the porous medium are not modeled with explicit coupling conditions, but taken into account via their interactions with the interface domain.

If a drop is sitting on a pore, only the liquid phase can flow across the interface through this pore.
Gas-filled pores without drops allow direct gas fluxes between free flow and porous medium.
The flux between free flow and interface consists of direct gas fluxes coming from the porous medium as well as evaporative fluxes from the drops on liquid-filled pores.

For local thermodynamic equilibrium, only the subscripts in Fquations (\ref{eq:chem-equil}) and (\ref{eq:therm-equil}) are adapted such that $\left[\dots \right]^\text{ff} = \left[\dots \right]^\text{if}$ and $\left[\dots \right]^\text{pm} = \left[\dots \right]^\text{if}$.
The equations for mechanical equilibrium are given as
\begin{align}
 \left[ \textbf{n} \cdot \left( \left(\varrho_g \textbf{v}_g \textbf{v}_g - \mu_g (\nabla \textbf{v}_g + \textbf{v}_g^T) + p_g \textbf{I} \right) \textbf{n} \right) \right]^{\text{ff}} &= \left[ p_g \right]^{\text{if}}\;, \nonumber \\
 \left[p_g \right]^{\text{if}} &= \left[p_g \right]^{\text{pm}}\;.
\end{align}

The Beavers-Joseph-Saffman condition (\ref{eq:BJS}) is now only applied to the free-flow/interface coupling.

For the continuity of fluxes, the area fractions presented in Section \ref{sec:homog} are used.
The flux between the porous medium and the interface is then given as
\begin{equation}
 q^{\text{pm}} = [(\varrho_g \textbf{v}_g) \cdot \textbf{n}]^{\text{up}} a_g A_{\Gamma}
  + [(\varrho_l \textbf{v}_l) \cdot \textbf{n}]^{\text{up}} a_l A_{\Gamma}\;,
  \label{eq:qpm}
\end{equation}
where $\varrho_{\alpha}$ and $\textbf{v}_{\alpha}$ are taken from the upstream domain
 ${\text{up}} \in \{\text{pm}, \text{if} \}$.
 Between free flow and interface domain,
 only the gas phase can be exchanged:
 \begin{equation}
  q^{\text{ff}} = [(\varrho_g \textbf{v}_g) \cdot \textbf{n}]^{\text{up}}
  \underbrace{(a_g + a_l)}_{=1} A_{\Gamma}\;,
  \label{eq:qff}
 \end{equation}
  with ${\text{up}} \in \{\text{ff}, \text{if} \}$.\\

The same applies to the component mass and energy fluxes.

\section{Numerical model}
\label{sec:numerical}
\begin{figure}
 \centering
 \includegraphics[width=0.35\textwidth]{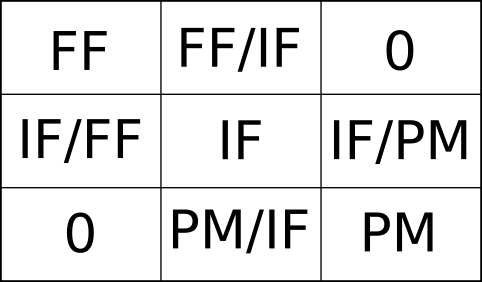}
 \caption{Structure of the global Jacobian matrix for the three-domain coupled system.}
 \label{fig:matrix}
\end{figure}

Discretizing the balance equations in Sections \ref{sec:ff}, \ref{sec:pm} and \ref{sec:if}
as well as the coupling conditions in Section \ref{sec:dropcoupling}
leads to a global non-linear system
\begin{equation}
 \textbf{J}(\textbf{u}) \cdot \textbf{u} = \textbf{b},
\end{equation}
where $\textbf{J}$ is the Jacobian matrix, $\textbf{u}$ the vector of unknowns
and $\textbf{b}$ the right-hand side.
The whole system is solved monolithically with the Newton's method.
The structure of the global Jacobian matrix $\textbf{J}$ is shown in figure \ref{fig:matrix}.
It contains three sub-matrices for the three domains on the diagonal
as well as four coupling matrices on the off-diagonals.
The lower left and upper right sub-matrices are zero matrices because the free flow and the porous medium
do not interact directly.

\subsection{Discretization}
\begin{figure}
 \begin{subfigure}{0.2\textwidth}
  \includegraphics[width=\linewidth]{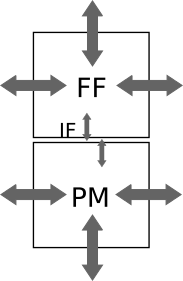}
  \caption{Without lateral fluxes}
  \label{fig:interaction_vertical}
 \end{subfigure}
 \hspace*{\fill}
 \begin{subfigure}{0.2\textwidth}
  \includegraphics[width=\linewidth]{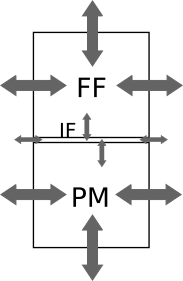}
  \caption{With lateral fluxes}
  \label{fig:interaction_horizontal}
 \end{subfigure}
 \caption{Interaction between interface grid cells only happens if lateral fluxes along the interface are taken into account}
 \label{fig:interaction}
\end{figure}

For the temporal discretization, a fully-implicit Euler scheme is applied.
The spatial discretizations are based on finite-volume methods.
We use a staggered grid \cite{Harlow1965} in the free flow domain.
Shifting the velocity components by half a cell to the cell boundaries
avoids pressure oscillations caused by numerics.
The interface domain and the porous medium domain are discretized
with cell-centered finite volumes and a two-point flux approximation.
For now, only square grid cells which coincide in conformity are used. \\
For the interaction regions, two scenarios have to be distinguished.
If lateral fluxes along the interface are neglected, no exchange between interface grid cells takes place, as shown in Figure \ref{fig:interaction_vertical}.
Taking lateral fluxes into account leads to a full interaction scheme between all neighboring cells, see Figure \ref{fig:interaction_horizontal}.

\subsection{Lower-dimensional interface domain}
The lower-dimensional interface domain $\Gamma$ is defined by the intersections of neighboring free-flow and porous-medium-flow grid cells.
This approach is based on \cite{Glaeser2017}, where fractures are modeled as lower-dimensional interface domains in a full-dimensional porous medium.
Within the interface domain, the balance equations (\ref{eq:if-mass}), (\ref{eq:if-comp}) and (\ref{eq:if-energy})
are integrated over the respective grid cells $\Omega_i$:
\begin{equation}
 \int_{\Omega_i} \frac{\partial u}{\partial t} + \nabla \cdot \textbf{F}(u)~d\Omega_i = \int_{\Omega_i} q_u~d\Omega_i\;,
 \label{eq:integ}
\end{equation}
where $u$ is either the pressure, momentum or energy respectively.
The coupling fluxes $q^{\text{ff}}$ and $q^{\text{pm}}$ are implemented as sources in the interface domain.
In the two full-dimensional domains, Neumann boundary conditions serve as coupling conditions.

\subsection{Extrusion}
\begin{figure}
 \centering
 \includegraphics[width=0.25\textwidth]{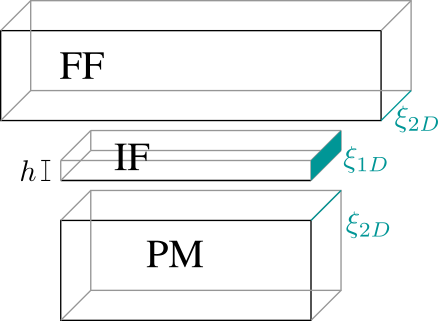}
 \caption{Extrusion to 3D according to the factors $\xi_1$ and $\xi_2$}
 \label{fig:extrusions}
\end{figure}

Evaluating the integrals as given in Equation (\ref{eq:integ}) would lead to inconsistent units for the full- and lower-dimensional domains.
Therefore, when computing volume-related terms in the lower-dimensional domain, the grid cell volume is multiplied by an extrusion factor $\xi_1$ in $m^2$
to extrude the one-dimensional to a three-dimensional domain.
All volume-related terms in the two-dimensional subdomains $\Omega^{\text{ff}}$ and
$\Omega^{\text{pm}}$ are multiplied by a factor $\xi_2$ in $m$ to extrude the domain in the third dimension.
This procedure is illustrated in Figure \ref{fig:extrusions}.
The lower-dimensional extrusion factor is chosen as $\xi_1 = \xi_2 \cdot h$, where $h$ is the maximum possible drop height for a drop sitting on the
largest pore.

\subsection{Pressure gradients}
\begin{figure}
 \centering
 \includegraphics[width=0.35\textwidth]{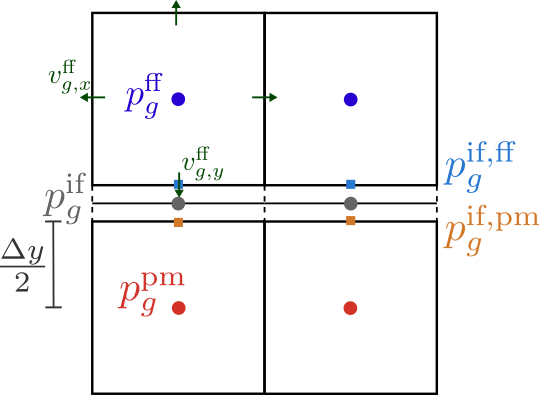}
 \caption{Grid positions of the actual (circles) and extrapolated (squares) gas pressures as well as the free-flow velocity components}
 \label{fig:pressure-grid}
\end{figure}

The pressure gradients to compute the convective fluxes across the interface depend on acutal and extrapolated pressure values as shown in Figure \ref{fig:pressure-grid}.
The circles represent actual pressures, which are computed and stored in the cell centers of the respective grid cells.
The squares represent extrapolated pressure values.
The gradients between actual and extrapolated values determine the fluxes from cell centers to cell faces.
The gas flux from the porous medium towards the interface depends on the pressure gradient between $p_g^{\text{pm}}$ and $p_g^{\text{if,pm}}$, which is represented by the lower orange squares.
The gradient between $p_g^{\text{if,pm}}$ and $p_g^{\text{if}}$ determines the gas flux from the interface domain's boundary towards its center.
On the other side, the blue squares represent the extrapolated pressure $p_g^{\text{if,ff}}$.
If a drop is present in an interface grid cell, the two gradients between $p_l^{\text{pm}}$ and $p_l^{\text{if,pm}}$ as well as between $p_l^{\text{if,pm}}$ and $p_l^{\text{if}}$
are computed accordingly.

Since the coupling flux $q^{\text{pm}}$ is computed between the cell centers of the respective domains, the extrapolated pressure values are eliminated.
Therefore, $q^{\text{pm}}$ can be computed with cell-center values only as given in Equation (\ref{eq:qpm}).
For the coupling flux $q^{\text{ff}}$, the velocity $v_{g,y}^{\text{ff}}$ is known on the grid face due to the staggered grid discretization and can be used directly.

\subsection{Algorithm}
Updating the drop information and solving the global system is done explicitely in two steps.
First, the drop volumes and area fractions are updated according to the initial solution or the solution of the previous time step.
Then, the global system is solved monolithically, applying the updated area fractions in the coupling conditions (\ref{eq:qpm}) and (\ref{eq:qff}).\\
The global system is solved monolithically with a Newton solver.
The linearized equations are then solved by the direct linear solver UMFPack \cite{DavisUMFPack}.

\section{Results}
\label{sec:results}
The coupled model is implemented in DuMu$^x$ (\cite{Dumux}, \cite{Dumux3}), an open-source simulator for flow in porous media and free-flow scenarios.
The code to reproduce the results of this work is available under
https://git.iws.uni-stuttgart.de/dumux-pub/Ackermann2020b.

\begin{figure}
 \centering
 \includegraphics[width=0.5\textwidth]{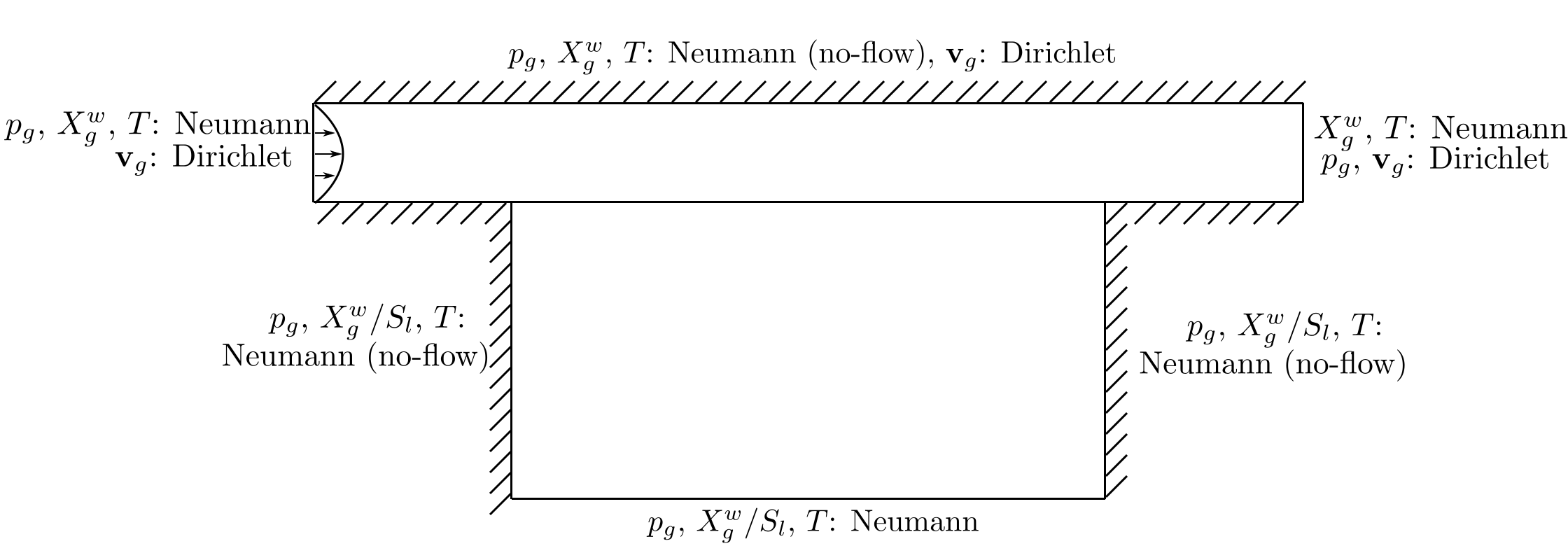}
 \caption{Model domain $\Omega$ with outer boundary conditions.}
 \label{fig:domain}
\end{figure}

The general set-up and the outer boundary conditions for all numerical experiments are depicted in Figure \ref{fig:domain}.
The free-flow domain extends the interface and porous-medium domains to ensure a stable velocity field.

\subsection{Drop formation, growth and detachment}
In the first numerical experiment, lateral fluxes along the interface are neglected.
Therefore, only storage and sink terms are evaluated in the interface domain $\Gamma$.
This means that only vertical fluxes between free flow and porous medium are taken into account.
For the sake of simplicity, only one pore-size class is assumed here.

\subsubsection{Initial and boundary values}
Initially, the gas pressure is set as $p_g = 10^5$Pa and the water mass fraction as $X_g^w = 0.01$ in the whole model domain $\Omega$.
A parabolic velocity profile is set for the free-flow velocity $\textbf{v}_g$ on the left boundary with a maximum of $v_{g,x} = 4.0\frac{m}{s}$.
The porous medium is initially completely gas-filled.
At the bottom boundary of the porous medium, an inflow rate of $q_{\text{bottom}} = 0.02 \frac{kg}{m^2 s}$ is set as a Neumann boundary condition.
Inflow and outflow conditions at the left and right boundaries of the free-flow domain ensure a continuous component intake and removal respectively.
If not stated differently, isothermal conditions are assumed.

\subsubsection{Drops in the interface domain}
\begin{figure}
 \centering
 \includegraphics[width=0.5\textwidth]{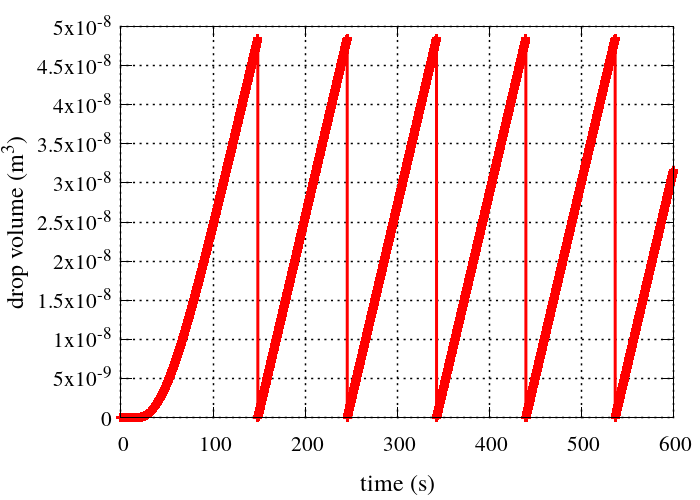}
 \caption{Drop volume over time: formation, growth and detachment}
 \label{fig:volumes_vertical}
\end{figure}

\begin{figure}
 \centering
 \includegraphics[width=0.5\textwidth]{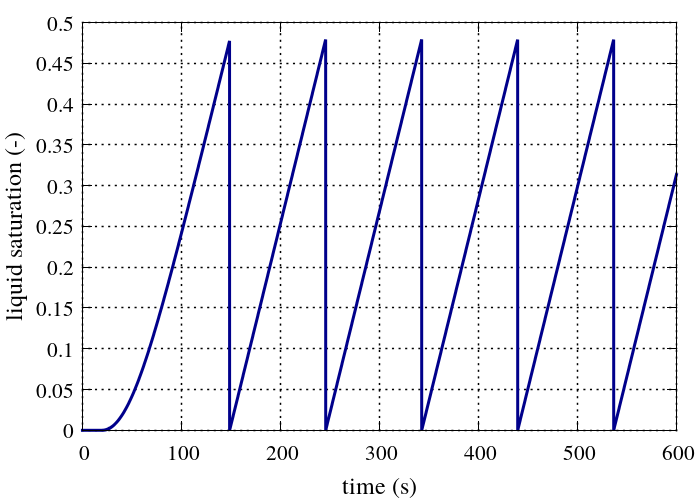}
 \caption{The liquid saturation evolution fits the drop volume evolution}
 \label{fig:saturation_vertical}
\end{figure}

Figure \ref{fig:volumes_vertical} shows the drop volume over time in a central interface element.
Due to the pressure gradient in the porous medium, the liquid phase reaches the interface
and drops form.
As soon as the drag force exceeds the retention force, the drops detach.
Due to the constant liquid inflow at the bottom boundary of the porous medium, liquid accumulates again which leads to a rising liquid pressure below the interface.
Since the formation condition (\ref{eq:formation}) is still fulfilled, a new drop forms and grows.
The same development can be observed for the liquid saturation in the interface domain in Figure \ref{fig:saturation_vertical}.
Due to the coupling fluxes, the primary variable $S_l$ increases according to the growing drop volume.
After the drop detachment, only the gas phase is left in the interface domain and $S_l = 0$.

\begin{figure}
 \centering
 \includegraphics[width=0.5\textwidth]{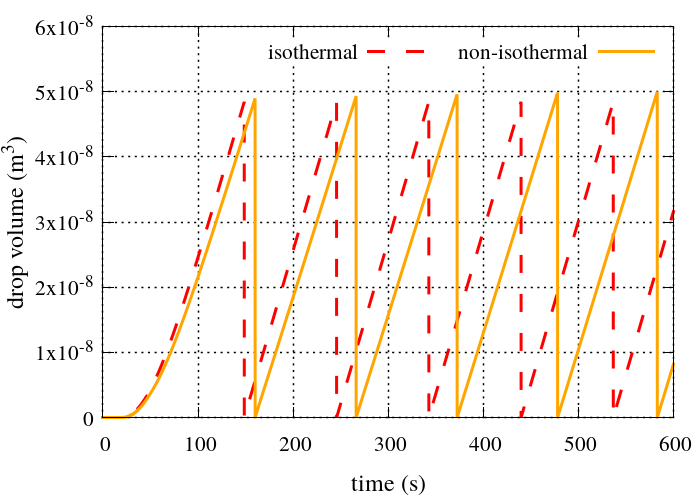}
 \caption{Drop volume evolution for isothermal and non-isothermal conditions}
 \label{fig:noniso}
\end{figure}

The growth of the drops is reduced under non-isothermal conditions, as shown in Figure \ref{fig:noniso}.
The evaporation of water vapor from the drops into the free flow slightly reduces the drop volume, leading to a delayed detachment compared to the isothermal case.

\subsubsection{Parameter variations}
\begin{figure}
 \centering
 \includegraphics[width=0.5\textwidth]{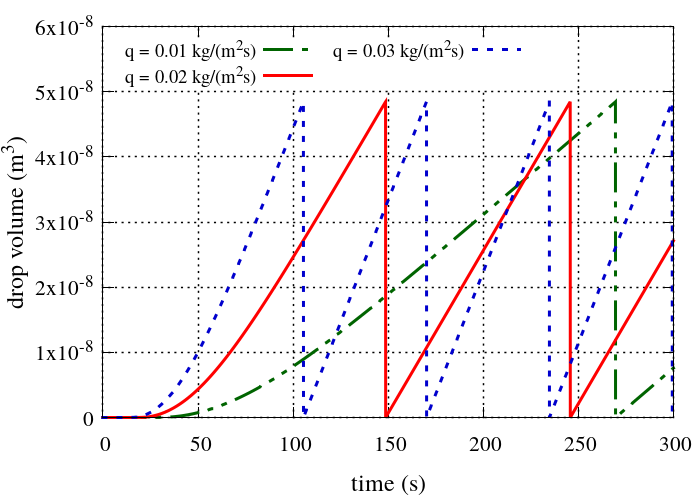}
 \caption{Drop volume evolution for varying inflow rates}
 \label{fig:inflow}
\end{figure}

The drop volume evolution depends on various parameters.
As shown in Figure \ref{fig:inflow}, reducing the inflow rate at the bottom boundary of the porous-medium domain leads to slower drop growth (green dashed line).
In contrast, a higher inflow rate accelerates the drop growth, leading to an earlier detachment compared to the reference case (red solid line).

Varying the free-flow velocity directly influences the drop detachment.
A higher velocity leads to a higher drag force, therefore the drops are detached earlier for $v_g^{\text{ff}} = 4.5 \frac{m}{s}$ (green dashed line) than for the reference case with $v_g^{\text{ff}} = 4 \frac{m}{s}$.
For smaller velocities, such as $v_g^{\text{ff}} = 3.5 \frac{m}{s}$ (blue dotted line), it takes longer for the drag force $F_{\text{drag}}$ to overcome the retention force $F_{\text{ret}}$, resulting in larger drop volumes at the time of detachment.

\begin{figure}
 \centering
 \includegraphics[width=0.5\textwidth]{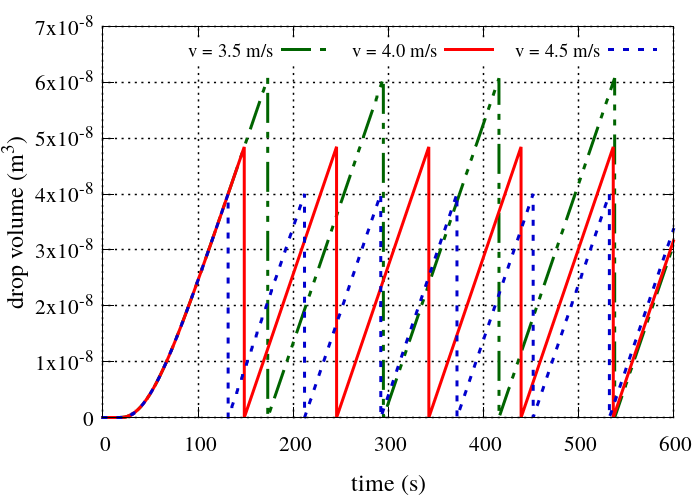}
 \caption{Drop volume evolution for varying free-flow velocities}
 \label{fig:velocity}
\end{figure}

\subsubsection{Influence on free flow and porous medium}
\begin{figure}
 \centering
 \includegraphics[width=0.4\textwidth]{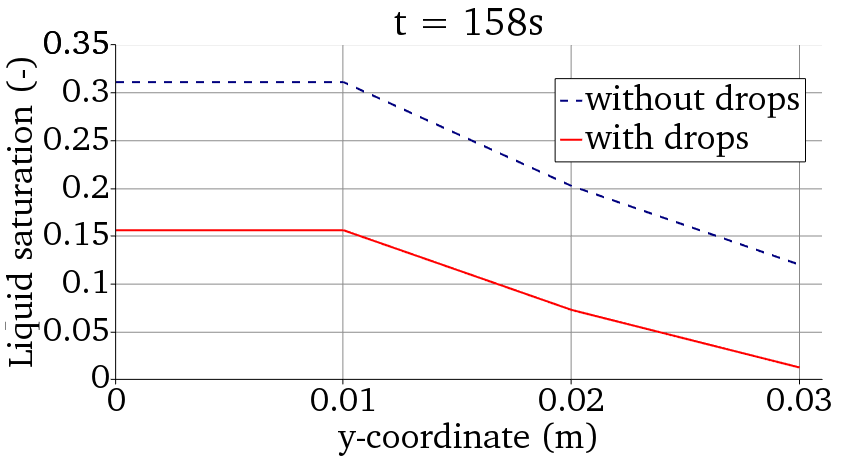}
 \caption{Liquid saturation in the porous medium}
 \label{fig:pm-Sl}
\end{figure}

Besides the drop evolution at the interface, the drops' influence on the adjacent flow regimes can be captured.
Figure \ref{fig:pm-Sl} compares the resulting liquid saturation along the central y-axis of the porous medium for coupling without drops and coupling with drops.
If drops are neglected, as in the simple coupling concept, the saturation at the upper boundary of the porous medium rises continuously (dashed line).
Considering drops allows the liquid phase to leave porous medium across the upper boundary, leading to a reduced saturation (solid line).

\begin{figure}
 \centering
 \includegraphics[width=0.4\textwidth]{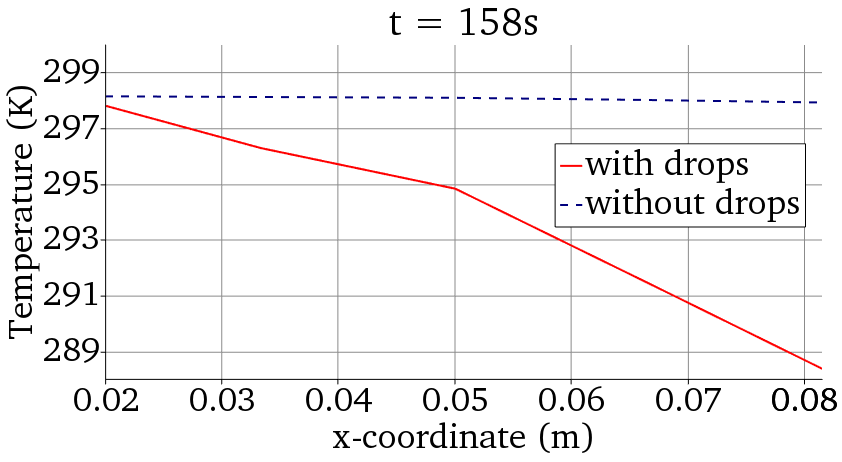}
 \caption{Temperature in the free flow}
 \label{fig:ff-T}
\end{figure}

\begin{figure}
 \centering
 \includegraphics[width=0.4\textwidth]{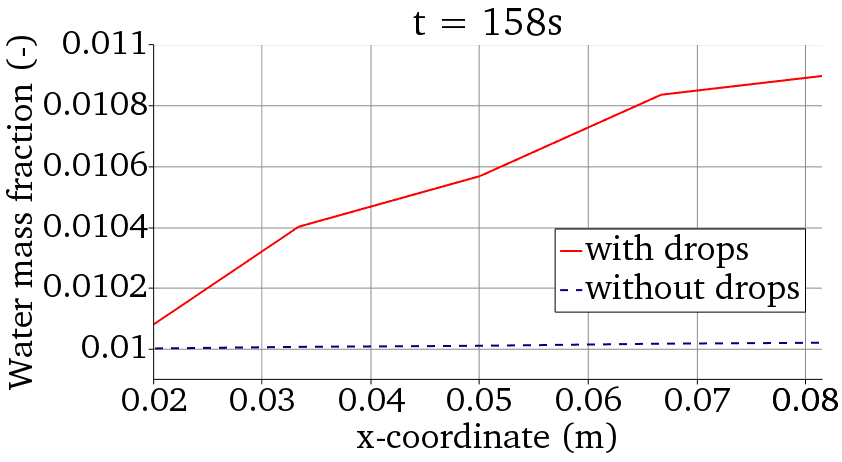}
 \caption{Water mass fraction in the free flow}
 \label{fig:ff-Xgw}
\end{figure}

Under non-isothermal conditions, the influence of the drops on the temperature $T$ and the mass fraction $X_g^w$ due to evaporation can be observed.
If drops are neglected, the temperature along th x-axis stays almost constant (dashed line), see Figure \ref{fig:ff-T}.
A coupling concept with drops leads to higher evaporative cooling, due to the larger liquid surface areas of the curved drops.
As a consequence, the water mass fraction in gas rises due to higher evaporation rates if drops are taken into account, compared to the result obtained with a coupling concept without drops, as shown in Figure \ref{fig:ff-Xgw}.

\subsection{Horizontal fluxes}
For the second numerical experiment, lateral fluxes along the interface are taken into account.
The gas phase as well as the liquid phase can now transfer mass, momentum and energy across
the boundaries of the interface grid cells.
For the free flow, the velocity is now set to zero, and a pressure gradient with a horizontal pressure difference of $\Delta p_g = 0.006$Pa is set.
On the right boundary of the interface domain, an outflow condition is set.

\begin{figure}
 \centering
 \includegraphics[width=0.5\textwidth]{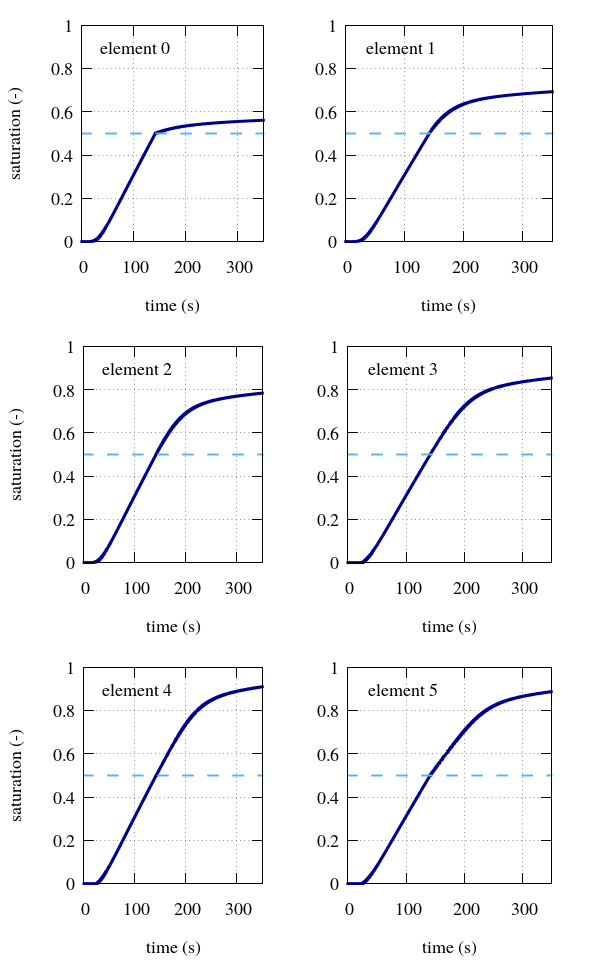}
 \caption{Saturation evolution due to lateral fluxes along the interface}
 \label{fig:lateral}
\end{figure}

Figure \ref{fig:lateral} shows the saturation evolution in each of the six interface grid cells.
We choose $S_l^{\text{merge}} = 0.5$ as the threshold saturation, where the drops are assumed to touch and merge.
Due to the horizontal pressure gradient, the liquid starts to flow from left to right, as soon as the threshold saturation is exceeded in two neighboring grid cells.
The water saturation is slightly higher in the right part of the interface domain,
but does not accumulate continuously due to the outflow condition on the right boundary.
After $t \approx 200$s, the saturation levels stay constant.
The saturation jumps between the grid cells are mainly caused by the coarse spatial discretization.

\section{Conclusions}
\label{sec:conclusions}
We present a new multi-scale coupling concept based on a three-domain approach to model drops at the interface of a free flow and a flow in a porous medium.
Criteria for drop formation, detachment and merging are formulated and evaluated to determine the drop volume evolution.
Additionally, the drops' presence at the interface is taken into account when computing the coupling fluxes between free flow and porous medium.

The results presented in the previous section show that drop formation, growth and detachment can be captured with the multi-scale coupling concept.
In addition, evaporation from the drops influences the drop volume evolution in case of non-isothermal conditions.
Since the liquid phase leaves the porous medium across its upper boundary, the liquid saturation is smaller compared to the one obtained with a simpler coupling concept without drops.
If drops are taken into account, the liquid surface area available for evaporative fluxes is larger than in the case without drops.
This leads to lower temperatures and higher water mass fractions in the free flow when drops are considered.

To obtain more precise results, the merging criterion should be determined with the help of experiments or analytical considerations.
Additionally, the influence of the drops on the free-flow velocity field should be taken into account.

\subsection*{Acknowledgments}
We would like to thank the Deutsche Forschungsgemeinschaft (DFG) for the financial support for this work in the frame of the International Research Training Group "Droplet Interaction Technologies" (DROPIT) and the SFB 1313, Project Number 327154368.

\appendix
\section{Derivation of relative permeabilities for interface domain}
\label{sec:appendix}
We here derive the relative permeabilities (\ref{eq:krg}) and (\ref{eq:krl}) from Section \ref{subsec:filmflow}, to describe the thin film and gas flow in the interface domain for the setting depicted in Figure \ref{fig:filmflow-domain}. Here we consider the full-dimensional description of $\Omega^{\text{if}}$, which is hence a layer of height $h$. In the layer there is a liquid film of height $0\leq \ell\leq h$. In the general case, $\ell=\ell(t,x,y)$. The interface domain is assumed to have length/width $L$, where $h<<L$ as the interface is thin. We divide $\Omega^{\text{if}}$ into the two time-dependent domains $\Omega_l^{\text{if}}(t)$ and $\Omega_g^{\text{if}}(t)$, with the separating interface $\Gamma^{\text{if}}(t)$:
\begin{align*}
\Omega_l^{\text{if}}(t) &= \{ 0< x,y< L,\ 0< z< \ell(t,x,y) \}, \\
\Omega_g^{\text{if}}(t) &= \{ 0< x,y< L,\ \ell(t,x,y)< z< h \}, \\
\Gamma^{\text{if}}(t) &= \{ 0< x,y< L,\  z= \ell(t,x,y) \}.
\end{align*}
We consider conservation of mass and momentum balance using Navier-Stokes for both fluids; namely (\ref{eq:ff-mass}) and (\ref{eq:ff-NS}) for gas in $\Omega_g^{\text{if}}(t)$, and 
\begin{align}
&\frac{\partial \varrho_l}{\partial t} + \nabla \cdot (\varrho_l \textbf{v}_l) = 0,\\
\nonumber
& \frac{\partial (\varrho_l \textbf{v}_l)}{\partial t} + \nabla \cdot \left(\varrho_l \textbf{v}_l \textbf{v}_l^{\text{T}} \right)\\ &= \nabla \cdot \left( \mu_l \left(\nabla \textbf{v}_l + \nabla \textbf{v}_l^T \right) - p_l \textbf{I} \right) + \varrho_l \textbf{g}\;.\label{eq:lig-NS}
\end{align}
for the water in $\Omega_l^{\text{if}}(t)$. At the interface $\Gamma^{\text{if}}(t)$ we have jump conditions ensuring conservation of mass and momentum:
\begin{align*}
(\varrho_l\mathbf v_l-\varrho_g\mathbf v_g)\cdot\mathbf n &= (\varrho_l-\varrho_g)v_n, \\
\mathbf v_l\cdot\mathbf t_i &= \mathbf v_g\cdot\mathbf t_i, \ i=1,2, \\
\big(-(p_l-p_g)\mathbf I + (\boldsymbol{\tau}_l-\boldsymbol{\tau}_g)\big)\cdot\mathbf n &= \varrho_l(\mathbf v_l\cdot\mathbf n-v_n)(\mathbf v_l-\mathbf v_g)\\&-\gamma_{lg}\kappa\mathbf n,
\end{align*}
where $\boldsymbol{\tau}_\alpha=\mu_\alpha(\nabla\mathbf v_\alpha+\nabla\mathbf v_\alpha^T)$. Further, the interface normal vector $\mathbf n$, tangent vectors $\mathbf t_1,\mathbf t_2$ and normal velocity $v_n$ are given by
\begin{align*}
\mathbf n &= \frac{(-\partial_x \ell,-\partial_y\ell,1)}{\sqrt{1+|\nabla \ell|^2}}, \quad v_n = \frac{\partial_t \ell}{\sqrt{1+|\nabla \ell|^2}},\\ \quad \mathbf t_1 &= \frac{(1,0,\partial_x \ell)}{\sqrt{1+(\partial_x \ell)^2}},\quad \mathbf t_2 = \frac{(0,1,\partial_y \ell)}{\sqrt{1+(\partial_y \ell)^2}},\\ 
\end{align*}
The curvature $\kappa$ is given by $\kappa = \nabla\cdot\mathbf n$. To derive the relative permeabilities in $\Omega^{\text{if}}$ we need boundary conditions for the top and bottom. We will for simplicity use
\begin{align*}
\mathbf v_g = \mathbf 0 \text{ on } z=h,\quad 
\mathbf v_l = \mathbf 0 \text{ on } z=0.
\end{align*}

We non-dimensionalize the equation to better determine which are important for the effective behavior. To this aim we introduce the small number $\varepsilon=h/L<<1$. We introduce the following, non-dimensional variables
\begin{align*}
\hat x,\hat y = \frac{x,y}{L},\quad \hat z = \frac{z}{h}, \quad \hat \ell = \frac{\ell}{h}, \quad \hat t = \frac{t}{t_{\text{ref}}}, \\ \hat{\mathbf v}_\alpha = \frac{\mathbf v_\alpha}{v_{\text{ref}}}, \quad \hat p_\alpha = \frac{p_{\alpha}}{p_{\text{ref}}}, \quad \hat \varrho_\alpha = \frac{\varrho_{\alpha}}{\varrho_{\text{ref}}}.
\end{align*}
We then have the non-dimensional domains and interface
\begin{align*}
\hat\Omega_l^{\text{if}}(\hat t) &= \{ 0< \hat x,\hat y<1, 0< \hat z< \hat \ell(\hat t,\hat x,\hat y) \}, \\
\hat\Omega_g^{\text{if}}(\hat t) &= \{0< \hat x,\hat y< 1, \hat \ell(\hat t,\hat x,\hat y)<\hat z< 1 \}, \\
\hat\Gamma^{\text{if}}(\hat t) &= \{ 0<\hat x,\hat y< 1,  \hat z= \hat \ell(\hat t,\hat x,\hat y) \},
\end{align*}
To remain in the range of Darcy's law we assume
\begin{align*}
\frac{\rho_{\text{ref}}Lv_{\text{ref}}}{\mu_l}=1,\quad \frac{p_{\text{ref}}}{v_{\text{ref}}^2\rho_{\text{ref}}}=\varepsilon^{-2}.
\end{align*}
We let $t_{\text{ref}}=L/v_{\text{ref}}$. As surface tension driven motion is not expected to be important for the thin film we assume $\mu_lv_{\text{ref}}/\gamma_{lg}=1$. For simplicity we assume that the viscosities are constant and introduce the viscosity ratio $M=\mu_l/\mu_g$. This gives the non-dimensional model equations and boundary conditions:
\begin{align}
&\frac{\partial\hat\varrho_\alpha}{\partial\hat t}+\hat\nabla\cdot(\hat\varrho_\alpha\hat{\mathbf v}_\alpha)= 0 \quad \text{ in } \hat \Omega_\alpha^{\text{if}}(\hat t), \label{eq:nondimmass} \\
&\varepsilon^2\Big(\frac{\partial (\hat\varrho_l\hat{\textbf{v}}_g)}{\partial t} + \hat\nabla \cdot \left(\hat\varrho_g \hat{\textbf{v}}_g \hat{\textbf{v}}_g^{\text{T}} \right)\Big)\nonumber= \varepsilon^2\frac{1}{M}\hat\nabla \cdot \left( \hat\nabla \hat{\textbf{v}}_g + \hat\nabla \hat{\textbf{v}}_g^T \right) \\&- \hat\nabla \hat p_g + \hat\varrho_g \hat{\textbf{g}} \quad \text{ in } \hat\Omega_g^{\text{if}}(t),\label{eq:nondimmomg}\\
&\varepsilon^2\Big(\frac{\partial (\hat\varrho_l\hat{\textbf{v}}_l)}{\partial t} + \hat\nabla \cdot \left(\hat\varrho_l \hat{\textbf{v}}_l \hat{\textbf{v}}_l^{\text{T}} \right)\Big)\nonumber= \varepsilon^2\hat\nabla \cdot \left( \hat\nabla \hat{\textbf{v}}_l + \hat\nabla \hat{\textbf{v}}_l^T \right) \\&- \hat\nabla \hat p_l + \hat\varrho_l \hat{\textbf{g}}\quad \text{ in } \hat\Omega_l^{\text{if}}(\hat t),\label{eq:nondimmoml}\\
&(\hat\varrho_l\hat{\mathbf v}_l-\hat\varrho_g\hat{\mathbf v}_g)\cdot\hat{\mathbf n}= (\hat\varrho_l-\hat\varrho_g)\hat v_n \quad \text{ on } \hat\Gamma^{\text{if}}(\hat t), \\
&\hat{\mathbf v}_l\cdot\hat{\mathbf t}_i = \hat{\mathbf v}_g\cdot\hat{\mathbf t}_i,\ i=1,2\quad \text{ on } \hat\Gamma^{\text{if}}(\hat t),\label{eq:nondimvelt} \\
&\big(-(\hat p_l-\hat p_g)\mathbf I + \varepsilon^2(\hat{\boldsymbol{\tau}}_l-\frac{1}{M}\hat{\boldsymbol{\tau}}_g)\big)\cdot\hat{\mathbf n}\nonumber\\&= \varepsilon^2(\hat{\mathbf v}_l\cdot\hat{\mathbf n}-\hat v_n)(\hat{\mathbf v}_l-\hat{\mathbf v}_g)-\varepsilon^2(\hat\nabla\cdot\hat{\mathbf n})\hat{\mathbf n}\ \text{  on } \hat\Gamma^{\text{if}}(\hat t),\label{eq:nondimmomgamma}\\
&\hat{\mathbf v}_l =\mathbf 0 \quad\text{ on } \hat z=0,\label{eq:nondimvlnoslip}\\
&\hat{\mathbf v}_g =\mathbf 0 \quad\text{ on } \hat z=1.\label{eq:nondimvgnoslip}
\end{align}
Here, the normal vector, tangent vector and normal velocity expressed using non-dimensional variables are
\begin{align*}
\hat{\mathbf n} &= \frac{(-\varepsilon\partial_{\hat x} \hat \ell,-\varepsilon\partial_{\hat y}\hat\ell,1)}{\sqrt{1+|\varepsilon\hat\nabla \hat\ell|^2}}, \quad \hat v_n = \frac{\varepsilon\partial_{\hat t}\hat\ell}{\sqrt{1+|\varepsilon\hat\nabla \hat\ell|^2}}, \\  \hat{\mathbf t}_1 &= \frac{(1,0,\varepsilon\partial_{\hat x} \hat\ell)}{\sqrt{1+(\varepsilon\partial_{\hat x} \hat\ell)^2}},\quad \hat{\mathbf t}_2 = \frac{(0,1,\varepsilon\partial_{\hat y} \hat\ell)}{\sqrt{1+(\varepsilon\partial_{\hat y} \hat\ell)^2}}.
\end{align*}
Further, $\hat{\boldsymbol{\tau}}=\hat\nabla\hat{\mathbf v}+\hat\nabla\hat{\mathbf v}^T$ and $\hat{\mathbf g} = L\mathbf g/v_{\text{ref}}^2$. However, we will for simplicity ignore the gravity as it does not affect the relative permeability. Finally, note that $\hat\nabla =(\partial_{\hat x},\partial_{\hat y},\frac{1}{\varepsilon}\partial_{\hat z})$.

We assume all variables have asymptotic expansions with respect to $\varepsilon$; that is,
\begin{align*}
\hat{\mathbf v}_\alpha = \hat{\mathbf v}_{\alpha,0}+\varepsilon\hat{\mathbf v}_{\alpha,1} + \varepsilon^2\hat{\mathbf v}_{\alpha,2}+\dots
\end{align*}
and similarly for the other variables. We insert these into (\ref{eq:nondimmass})-(\ref{eq:nondimvgnoslip}) and collect the dominating terms with respect to $\varepsilon$.

The dominating term from (\ref{eq:nondimmass}) gives
\begin{equation*}
\partial_{\hat z}\hat{\mathbf v}_{l,0}^{(3)} = 0 \text{ for } 0\leq \hat z\leq \hat\ell,\quad \partial_{\hat z}\hat{\mathbf v}_{g,0}^{(3)} = 0 \text{ for } \hat\ell\leq \hat z\leq 1,
\end{equation*}
which combined with the dominating terms from (\ref{eq:nondimvlnoslip}) and (\ref{eq:nondimvgnoslip}) leads to
\begin{equation*}
\hat{\mathbf v}_{l,0}^{(3)}\equiv 0 \text{ and } \hat{\mathbf v}_{g,0}^{(3)}\equiv 0 \text{ for all } \hat z.
\end{equation*}

Assuming that the viscosity ratio $M=O(1)$, we obtain from the dominating terms in (\ref{eq:nondimmomg}) and (\ref{eq:nondimmoml}) that
\begin{equation*}
\partial_{\hat z}\hat p_{\alpha,0} = 0,
\end{equation*}
which means that $\hat p_{\alpha,0}=\hat p_{\alpha,0}(\hat t,\hat x,\hat y)$.
Continuing to the next order terms, the first and second component give
\begin{align*}
\partial_{\hat z}^2\hat{\mathbf v}_{l,0}^{(1)} = \partial_{\hat x}\hat p_{l,0}\quad \text{and}\quad \partial_{\hat z}^2\hat{\mathbf v}_{g,0}^{(1)} = M\partial_{\hat x}\hat p_{g,0}, \\
\partial_{\hat z}^2\hat{\mathbf v}_{l,0}^{(2)} = \partial_{\hat y}\hat p_{l,0}\quad \text{and}\quad \partial_{\hat z}^2\hat{\mathbf v}_{g,0}^{(2)} = M\partial_{\hat y}\hat p_{g,0}.
\end{align*}
Integrating these twice with respect to $\hat z$, and using the dominating terms from (\ref{eq:nondimvelt}), (\ref{eq:nondimvlnoslip}) and (\ref{eq:nondimvgnoslip}) together with the $O(\varepsilon)$ from (\ref{eq:nondimmomgamma}) to determine the integration constants, we get
\begin{align*}
(\hat{\mathbf v}_{l,0}^{(1)},\hat{\mathbf v}_{l,0}^{(2)}) &= \frac{1}{2}(\partial_{\hat x}\hat p_{l,0},\partial_{\hat y}\hat p_{l,0})(\hat z^2 - \frac{M(1-\hat\ell_0^2)+\hat\ell_0^2}{M(1-\hat\ell_0)+\hat\ell_0}\hat z),\\
(\hat{\mathbf v}_{g,0}^{(1)},\hat{\mathbf v}_{g,0}^{(2)}) &= \frac{1}{2}M(\partial_{\hat x}\hat p_{g,0},\partial_{\hat y}\hat p_{g,0})\big((\hat z^2-1)\\ &- \frac{1}{2}M \frac{M(1-\hat\ell_0^2)+\hat\ell_0^2}{M(1-\hat\ell_0)+\hat\ell_0}(\hat z-1)\big).
\end{align*}
By averaging the lowest order velocities over the height of the domain (which in this case is 1) we obtain the (non-dimensional) Darcy velocities:
\begin{align*}
\hat{\overline{\mathbf v}}_{l,0} &= \int_0^{\hat\ell_0}\hat{\mathbf v}_{l,0}d\hat z\\ &= - \frac{1}{12}\hat\nabla \hat p_{l,0}\Big(-2\hat\ell_0^3 + 3\frac{M(1-\hat\ell_0^2)+\hat\ell_0^2}{M(1-\hat\ell_0)-\hat\ell_0}\hat\ell_0^2\Big), \\
\hat{\overline{\mathbf v}}_{g,0} &= \int_{\hat\ell_0}^1\hat{\mathbf v}_{g,0}d\hat z\\ &= - \frac{M}{12}\hat\nabla\hat p_{g,0}(\hat\ell_0-1)^2\Big(2(\hat\ell_0+2) - 3\frac{M(1-\hat\ell_0^2)+\hat\ell_0^2}{M(1-\hat\ell_0)-\hat\ell_0}\Big).
\end{align*}
By using the lowest order variables in the interface domain, we obtain when returning to dimensional variables
\begin{align*}
{\mathbf v}_{l}^{\text{if}} &= - \frac{\mathbf K}{\mu_l}S_l^2\Big(-2S_l + 3\frac{\mu_l(1-S_l^2)+\mu_gS^2}{\mu_l(1-S_l)+\mu_gS_l}\Big)\nabla p_{l}, \\
{\mathbf v}_{g}^{\text{if}} &= - \frac{\mathbf K}{\mu_g}(1-S_l)^2\Big(2(S_l+2) - 3\frac{\mu_l(1-S_l^2)+\mu_gS_l^2}{\mu_l(1-S_l)+\mu_gS_l}\Big)\nabla p_{g}.
\end{align*}
where we have used $S_l=\ell/h$ for the saturation and $\mathbf K=\frac{h^2}{12}\mathbf I$. Rewriting this way enables us to identify the relative permeabilities:
\begin{align}
k_{rl}(S_l) &= S_l^2\Big(-2S_l + 3\frac{\mu_l(1-S_l^2)+\mu_gS_l^2}{\mu_l(1-S_l)+\mu_gS_l}\Big),\label{eq:upscaledalt} \\
k_{rg}(S_l) &= (1-S_l)^2\Big(2(S_l+2) - 3\frac{\mu_l(1-S_l^2)+\mu_gS_l^2}{\mu_l(1-S_l)+\mu_gS_l}\Big).\label{eq:upscaledrelg}
\end{align}

Since the viscosity ratio between water and gas is quite large, an alternative relative permeability can be derived for the liquid phase. Letting $M=O(\varepsilon^{-1})$ and repeating the steps we obtain
\begin{equation*}
{\mathbf v}_{l}^{\text{if}} = -\frac{\mathbf K}{\mu_l}4S_l^3\nabla p_{l},
\end{equation*}
which corresponds to the relative permeability
\begin{equation}
k_{rl} (S_l) = 4S_l^3.\label{eq:upscaledrell}
\end{equation}
With this approach no expression is obtained for the velocity in the gas phase. We hence use (\ref{eq:upscaledrelg}) for the relative permeability of the gas phase, and (\ref{eq:upscaledrell}) for the liquid phase.

\bibliographystyle{model1-num-names}
\bibliography{refs}

\end{document}